\documentclass{ws-procs9x6}
\usepackage{amsmath}
\usepackage{graphicx}
\usepackage{amsfonts}
\usepackage{amssymb}
\begin{document}

\title{Plasma expansion in the geometry of a collapsing star}
\author{R. RUFFINI, L. VITAGLIANO AND S.-S. XUE}
\maketitle

\address{ICRA and
Physics Department, \\
University of Rome ``La Sapienza", \\
P.le A. Moro 5, \\
00185 Rome, Italy}

\abstracts
{We describe the evolution of an electron-positron-photon plasma created by Sauter--Heisenberg--Euler--Schwinger mechanism around
a collapsing charged star core in the Reissner-Nordstr\"{o}%
m geometry external to
the core, in view of the application in the framework of the EMBH theory for gamma ray bursts.}

\section{Introduction}

In 1975, following the work on the energetics of black holes,\cite{ruffc}
Damour and Ruffini\cite{DR75} pointed out the existence of the vacuum
polarization process \textit{\`{a} l\`{a}}
Sauter--Heisenberg--Euler--Schwinger\cite{he35,s51} around black holes endowed
with electromagnetic structure (EMBHs), whose electric field strength exceeds
the Schwinger critical value $\mathcal{E}_{\mathrm{c}}=\tfrac{m_{e}^{2}c^{3}%
}{e\hbar},$ where $c$ is the speed of light, $e$ and $m_{e}$ are electron
charge and mass respectively. Damour and Ruffini gave reasons to believe that
this process is almost reversible in the sense introduced by Christodoulou and
Ruffini\cite{ruffc} and that it extracts the mass energy of an EMBH very
efficiently: this have been proved in Ref.~\refcite{RV02}. The vacuum
polarization process around an EMBH offered a natural mechanism for explaining
the Gamma Ray Bursts (GRBs), just discovered at the time. Moreover the
mechanism had a most peculiar prediction: the characteristic energetics of the
burst should be of the order of $10^{54}$ ergs; while nothing at the time was
known about either the distances or the energetics of GRBs.

More recently, after the discovery of the afterglow of GRBs and their
cosmological distance, the idea by Damour and Ruffini has been reconsidered in
Refs.~\refcite{PRX98}--\refcite{RBCFX01c} where the EMBH model for GRBs is
developed. The evidence is now that through the observations of GRBs we are
witnessing the formation of an EMBH and therefore are following the process of
gravitational collapse in real time. Even more importantly, the tremendous
energies involved in the energetics of these sources have their origin in the
extractable energy of black holes.

Various models have been proposed in order to extract the rotational energy of
black holes by processes of relativistic magnetohydrodynamics (see, e.g.,
Ref.~\refcite{rw75}). It should be expected, however, that these processes are
relevant over the long time scales characteristic of accretion processes. In
the present case of GRBs a sudden mechanism appears to be at work on time
scales of the order of few seconds or shorter and they are naturally explained
by the vacuum polarization process introduced in Ref.~\refcite{DR75}.

All considerations on the electric charge of stars have been traditionally
directed towards the presence of a net charge on the star surface in a steady
state condition, from the classic work by Shvartsman\cite{s70} all the way to
the fundamental book by Punsly.\cite{punsly_book} The charge separation can
occur in stars endowed with rotation and magnetic field and surrounded by
plasma, as in the case of Goldreich and Julian,\cite{gj69} or in the case of
absence of both magnetic field and rotation, the electrostatic processes can
be related to the depth of the gravitational well, as in the treatment of
Shvartsman.\cite{s70} However, in neither case is it possible to reach the
condition of the overcritical field needed for pair creation.

\begin{figure}[ptb]
\vspace{-.5cm} \epsfxsize=\hsize
\par
\begin{center}
\mbox{\epsfbox{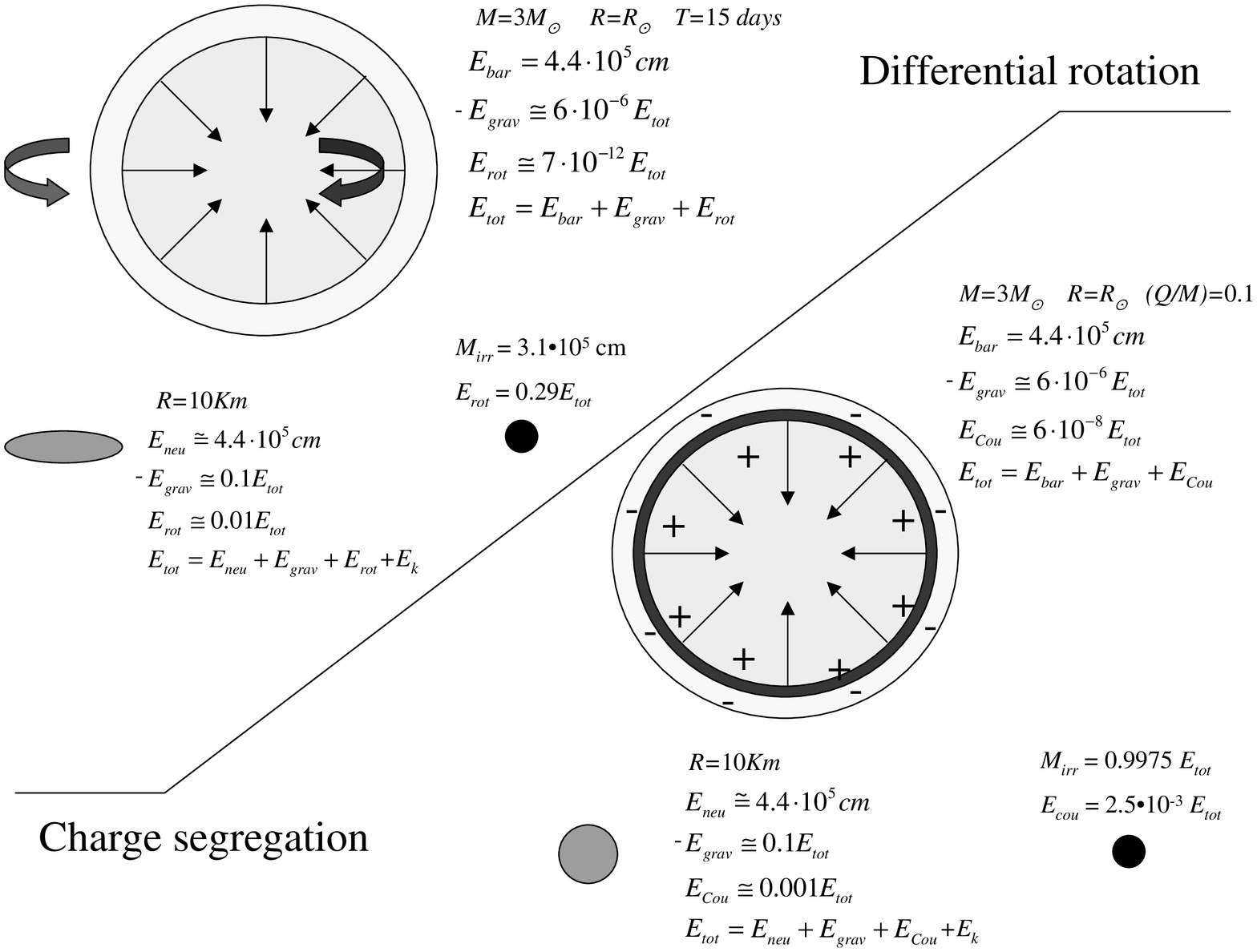}}
\end{center}
\par
\vspace{-0.2cm}\caption{Comparing and contrasting gravitational collapse to a
neutron star and to a black hole for a star core endowed with rotation or
electromagnetic structure. Reproduced from Ref.~[14] with the kind
permission of the author.}%
\label{twosph}%
\end{figure}

The basic new conceptual point is that GRBs are the most violent transient
phenomenon in the universe and therefore in order to realize the condition for
their occurrence, one must look at a transient phenomenon. We propose as a
candidate the most transient phenomenon possibly occurring in the life of a
star: the gravitational collapse. The condition for the creation of the
supercritical electromagnetic field required in the Damour and Ruffini work
has to be achieved during the process of gravitational collapse which lasts
less than $\sim30$ seconds for a mass of $10M_{\odot}$ and the relevant part
of the process may be as short as $10^{-2}$ or even $10^{-3}$ seconds. It is
appropriate to consider a numerical example here\cite{RMG} (see
Fig.~\ref{twosph}). We compare and contrast the gravitational collapse of a
star in the two limiting cases in which its core of $M=3M_{\odot}$ and radius
$R=R_{\odot}$ is either endowed with rotation or with electromagnetic
structure. The two possible outcomes of the process of gravitational collapse
are considered: either a neutron star of radius of $10\mathrm{km}$ or a black hole.

In the case of rotation the core has been assumed to have a rotational period
of $\sim15$ days. For such an initial configuration we have:
\begin{equation}
E_{rot}\simeq7\times10^{-12} E_{tot} \ll\left|  E_{grav}\right|  \simeq
6\times10^{-6} E_{tot} \ll E_{bar}\simeq4.4\times10^{5} \mathrm{cm}\, .
\label{rotcase1}%
\end{equation}
In the collapse to a neutron star we have:
\begin{equation}
E_{rot}\simeq0.01 E_{tot} \ll\left|  E_{grav}\right|  \simeq0.1 E_{tot} \ll
E_{bar}\simeq4.4\times10^{5} \mathrm{cm}\, . \label{rotcase2}%
\end{equation}
The very large increase in the rotational energy is clearly due to the process
of gravitational collapse: such a storage of rotational energy is the well
known process explaining the pulsar phenomena. The collapse to a black hole
has been estimated \emph{assuming} the mass--energy formula.\cite{ruffc} The
overall energetics, for the chosen set of parameters, leads to a solution
corresponding to an extreme black hole, for which in principle 29\% of the
energy is extractable.

The similar process in the electromagnetic case starts from an initial neutral
star with a magnetosphere oppositely charged from a core with
\begin{equation}
\tfrac{Q}{M\sqrt{G}}=0.1\;. \label{gc_eq1}%
\end{equation}
Let us first evaluate the amount of polarization needed in order to reach the
above relativistic condition. Recalling that the charge to mass ratio of a
proton is $q_{p}/\left(  m_{p}\sqrt{G}\right)  =1.1\times10^{18}$, it is
enough to have an excess of one quantum of charge every $10^{19}$ nucleons in
the core of the collapsing star to obtain such an EMBH after the occurrence of
the gravitational collapse. Physically this means that we are dealing with a
process of charge segregation between the core and the outer part of the star
which has the opposite sign of net charge in order to enforce the overall
charge neutrality condition.

We then have:
\begin{equation}
E_{cou}\simeq6\times10^{-8}E_{tot}\ll\left|  E_{grav}\right|  \simeq
6\times10^{-6}E_{tot}\ll E_{bar}\simeq4.4\times10^{5}\mathrm{cm}\,.
\label{emcase1}%
\end{equation}
In the collapse to the neutron star configuration we have:
\begin{equation}
E_{Cou}\simeq0.001E_{tot}\ll\left|  E_{grav}\right|  \simeq0.1E_{tot}\ll
E_{bar}\simeq4.4\times10^{5}\mathrm{cm}\,. \label{emcase2}%
\end{equation}
Once again, the amplification of the electromagnetic energy is due to the
process of gravitational collapse. Again, \emph{assuming} the mass--energy
formula, the collapse to a black hole for the chosen set of parameters leads
to:
\begin{equation}
M_{\mathrm{ir}}=0.9975E_{tot}\,,\quad E_{Cou}=2.5\times10^{-3}E_{tot}\,.
\label{emcase3}%
\end{equation}
It is during such a process of gravitational collapse to an EMBH that the
overcritical field is reached.

The process of charge segregation between the inner core and the oppositely
charged outer shell is likely due to the combined effects of rotation and
magnetic fields in the earliest phases of the gravitational collapse of the
progenitor star or to a process of ionization. In the following we will forget
about the outer shell and will treat the inner core as an electrically charged
collapsing star.

\section{Energy extraction from a supercritical EMBH}

We know from the Christodoulou-Ruffini mass formula\cite{ruffc} that the mass
energy of an EMBH is the sum of the irreducible mass and the electromagnetic
energy:
\begin{equation}
M=M_{\mathrm{ir}}+\tfrac{Q^{2}}{2r_{+}}, \label{irrmass}%
\end{equation}
where $Q$ is the charge and $r_{+}$ is the radius of the horizon. Moreover in
Ref.~\refcite{RV02} it is shown that the electromagnetic energy $\tfrac{Q^{2}%
}{2r_{+}}$ is stored throughout the region external to the EMBH and can be
extracted. If the condition
\begin{equation}
\tfrac{Q}{r_{+}^{2}}\geq\mathcal{E}_{\mathrm{c}} \label{condition}%
\end{equation}
is fulfilled the leading extraction process is a \emph{collective} process
based on the electron-positron plasma generated by Schwinger mechanism in the
supercritical electric field of the EMBH.\cite{RV02} The condition
(\ref{condition}) implies
\begin{equation}
\tfrac{GM/c^{2}}{\lambda_{\mathrm{C}}}\left(  \tfrac{e}{\sqrt{G}m_{e}}\right)
^{-1}\simeq2\cdot10^{-6}\tfrac{M}{M_{\odot}}\leq\xi\leq1\,
\end{equation}
and therefore this vacuum polarization process can occur only for an EMBH with
mass smaller than $2\cdot10^{6}M_{\odot}$. The electron-positron pairs are
produced in the dyadosphere of the EMBH,\cite{PRX98} the spherical region
whose radius $r_{\mathrm{ds}}$ satisfies $\mathcal{E}_{\mathrm{c}}\equiv
\tfrac{Q}{r_{\mathrm{ds}}^{2}}$. We have
\begin{equation}
r_{\mathrm{ds}}=\sqrt{\tfrac{eQ\hbar}{m_{e}^{2}c^{3}}}\,. \label{dya1}%
\end{equation}
The number of particles created is\cite{PRX98}
\begin{equation}
N_{\mathrm{ds}}=\tfrac{1}{3}\left(  \tfrac{r_{\mathrm{ds}}}{\lambda
_{\mathrm{C}}}\right)  \left(  1-\tfrac{r_{+}}{r_{\mathrm{ds}}}\right)
\left[  4+\tfrac{r_{+}}{r_{\mathrm{ds}}}+\left(  \tfrac{r_{+}}{r_{\mathrm{ds}%
}}\right)  ^{2}\right]  \tfrac{Q}{e}\simeq\tfrac{4}{3}\left(  \tfrac
{r_{\mathrm{ds}}}{\lambda_{\mathrm{C}}}\right)  \tfrac{Q}{e}\,. \label{numdya}%
\end{equation}
The total energy stored in the dyadosphere is\cite{PRX98}
\begin{equation}
E_{\mathrm{ds}}^{\mathrm{tot}}=\left(  1-\tfrac{r_{+}}{r_{\mathrm{ds}}%
}\right)  \left[  1-\left(  \tfrac{r_{+}}{r_{\mathrm{ds}}}\right)
^{4}\right]  \tfrac{Q^{2}}{2r_{+}}\simeq\tfrac{Q^{2}}{2r_{+}}\,.
\label{enedya}%
\end{equation}
The mean energy per particle produced in the dyadosphere $\left\langle
E\right\rangle _{\mathrm{ds}}=\tfrac{E_{\mathrm{ds}}^{\mathrm{tot}}%
}{N_{\mathrm{ds}}}$ is then
\begin{equation}
\left\langle E\right\rangle _{\mathrm{ds}}=\tfrac{3}{2}\tfrac{1-\left(
\tfrac{r_{+}}{r_{\mathrm{ds}}}\right)  ^{4}}{4+\tfrac{r_{+}}{r_{\mathrm{ds}}%
}+\left(  \tfrac{r_{+}}{r_{\mathrm{ds}}}\right)  ^{2}}\left(  \tfrac
{\lambda_{\mathrm{C}}}{r_{\mathrm{ds}}}\right)  \tfrac{Qe}{r_{+}}\simeq
\tfrac{3}{8}\left(  \tfrac{\lambda_{\mathrm{C}}}{r_{\mathrm{ds}}}\right)
\tfrac{Qe}{r_{+}}\,, \label{meanenedya}%
\end{equation}
which can be rewritten as
\begin{equation}
\left\langle E\right\rangle _{\mathrm{ds}}\simeq\tfrac{3}{8}\left(
\tfrac{r_{\mathrm{ds}}}{r_{+}}\right)  \ m_{e}c^{2}\sim\sqrt{\tfrac{\xi
}{M/M_{\odot}}}10^{5}keV\,. \label{GRB}%
\end{equation}
Such a process of vacuum polarization, occurring not at the horizon but in the
extended dyadosphere region ($r_{+}\leq r\leq r_{\mathrm{ds}}$) around an
EMBH, has been observed to reach the maximum efficiency limit of $50\%$ of the
total mass-energy for an extreme EMBH (see e.g.~Ref.~\refcite{PRX98}). As
discussed in Ref.~\refcite{RV02} the $e^{+}e^{-}$ creation process occurs at
the expense of the Coulomb energy and does not affect the irreducible mass,
which does not depend of the electromagnetic energy. In this sense, $\delta
M_{\mathrm{ir}}=0$ and the transformation is fully reversible.

\section{The EMBH Theory}

In a series of papers,\cite{PRX98}$^{-}$\cite{RBCFX01c} Ruffini and
collaborators have developed the EMBH theory for GRBs, which has the
advantage, despite its simplicity, that all eras following the process of
gravitational collapse to the EMBH are described by precise field equations
which can then be numerically integrated.

Starting from the vacuum polarization process \textit{\`{a} l\`{a}}
Sauter--Heisenberg--Euler--Schwinger in the overcritical field of an EMBH
first computed in Ref.~\refcite{DR75}, Ruffini et al. developed the
dyadosphere concept.\cite{PRX98}

The dynamics of the $e^{+}e^{-}$--pairs and electromagnetic radiation of the
plasma generated in the dyadosphere propagating away from the EMBH in a sharp
pulse (PEM pulse) has been studied by the Rome group and validated by the
numerical codes developed at Livermore Lab.\cite{RSWX99,RSWX00}

The collision of the still optically thick $e^{+}e^{-}$--pairs and
electromagnetic radiation plasma with the baryonic matter of the remnant of
the progenitor star has been again studied by the Rome group and validated by
the Livermore Lab codes.\cite{RSWX99,RSWX00} The further evolution of the
sharp pulse of pairs, electromagnetic radiation and baryons (PEMB pulse) has
been followed for increasing values of the gamma factor until the condition of
transparency is reached.\cite{BRX00}

\begin{figure}[ptb]
\begin{center}
\includegraphics[width=8.5cm,clip]{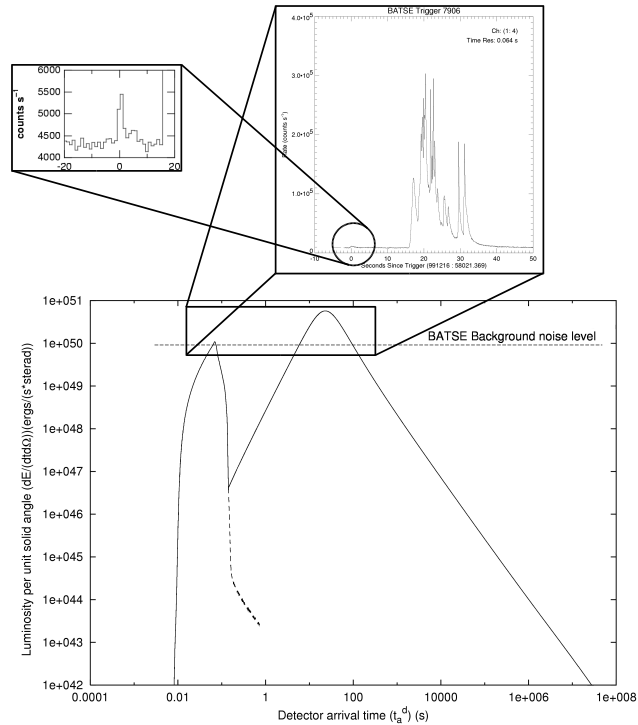}
\end{center}
\caption{The overall description of the EMBH theory applied to GRB~991216. The
BATSE noise threshold is represented and the observations both of the P-GRB
and of the E-APE are clearly shown in the subpanels. The continuos line in the
picture represents the theoretical prediction of the EMBH model.}%
\label{fit}%
\end{figure}

As this PEMB pulse reaches transparency the proper GRB (P-GRB) is
emitted\cite{RBCFX01b} and a pulse of accelerated baryonic matter (the ABM
pulse) is injected into the interstellar medium (ISM) giving rise to an
afterglow. Thus{ in GRBs we can distinguish an injector phase and a
beam-target phase. The injector phase includes the process of gravitational
collapse, the formation of the dyadosphere, as well as the PEM pulse, the
engulfment of the baryonic matter of the remnant and the PEMB pulse. The
injector phase terminates with the P-GRB emission. The beam-target phase
addresses the interaction of the ABM pulse, namely the beam generated during
the injection phase, with the ISM as the target. It gives rise to the E-APE
(Extended Afterglow Peak Emission) and the decaying part of the afterglow}.
The existence of both the P-GRB and the E-Ape is shown in Fig.~\ref{fit},
where the fit of observational data relative to GRB 991216 within the EMBH
theory is reported.

\section{Gravitational Collapse of an Electrically Charged Core: Formation of Dyadosphere}

We now turn to the details of the formation of dyadosphere. If the electric
field of a charged star core is stable against vacuum polarization during the
gravitational collapse,\cite{RVX03} then an enormous amount of pairs can be
created by Schwinger mechanism. Moreover the pairs thermalize to a
positrons-electrons-photons plasma configuration (see Refs.~\refcite{PRX98},
\refcite{RVX03a}, \refcite{RVX02}). Such a plasma undergoes a relativistic
expansion. The evolution of the system and the details of GRB emission, along
the lines summarized in the previous section, were described in Refs.~\refcite
{PRX98}--\refcite{RBCFX01b}, \refcite{RSWX99}--\refcite{BRX00}. In the latter
papers the time scale of the gravitational collapse is neglected with respect
to the hydrodynamic time scale. In this paper we relax this approximation: our
main aim is to describe how the plasma expansion occurs \emph{during} the
gravitational collapse. In a forthcoming paper\cite{RVX03} we will discuss how
the expansion is affected by the strong gravitational field near the horizon
of the forming EMBH.

In Refs.~\refcite{RVX02} and \refcite{CRV02} it was suggested that the exact
solution of Einstein-Maxwell equations describing the gravitational collapse
of a thin charged shell can be used as an analytical model for the
gravitational collapse of a charged core. First we briefly review some of the
results of Ref.~\refcite{CRV02}. The region of space-time external to the
collasping core is Reissner-Nordstr\"{o}m with line element, in Schwarzschild
like coordinates $\left(  t,r,\theta,\phi\right)  ,$
\begin{equation}
ds^{2}=-\alpha^{2}dt^{2}+\alpha^{-2}dr^{2}+r^{2}d\Omega^{2}, \label{ds}%
\end{equation}
where $d\Omega^{2}=d\theta^{2}+\sin^{2}\theta d\phi^{2}$, $\alpha^{2}%
=\alpha^{2}\left(  r\right)  =1-\tfrac{2M}{r}+\tfrac{Q^{2}}{r^{2}}$, $M$ is
the total energy of the core as measured at infinity and $Q$ is its total
charge. Let us label with $r_{0}$ and $t_{0}$ the radial and time-like
coordinates of the shell, then the electromagnetic field strength on the
surface of the core is $\mathcal{E}=\mathcal{E}\left(  r_{0}\right)  =\frac
{Q}{r_{0}^{2}}$ and the equation of core's collapse is\cite{CRV02}%

\begin{equation}
\tfrac{dr_{0}}{dt_{0}}=-\tfrac{\alpha^{2}\left(  r_{0}\right)  }{H\left(
r_{0}\right)  }\sqrt{H^{2}\left(  r_{0}\right)  -\alpha^{2}\left(
r_{0}\right)  } \label{Motion}%
\end{equation}
where $H\left(  r_{0}\right)  =\tfrac{M}{M_{0}}-\tfrac{M_{0}^{2}+Q^{2}}%
{2M_{0}r_{0}}$ and $M_{0}$ is the rest mass of the shell. The analytical
solution of Eq.~(\ref{Motion}) was found in Ref.~\refcite{CRV02} in the form
\begin{equation}
t_{0}=t_{0}\left(  r_{0}\right)  .
\end{equation}
Dyadosphere is formed since the instant $t_{\mathrm{ds}}=t_{0}\left(
r_{\mathrm{ds}}\right)  $ when $\mathcal{E}=\mathcal{E}_{\mathrm{c}}$. In the
following we put $t_{\mathrm{ds}}=0$.

\section{Formation of $e^{+}e^{-}$ Pairs around a Collapsing Charged Core}

For $t<$ $t_{\mathrm{ds}}$, $\mathcal{E}<\mathcal{E}_{\mathrm{c}}$ and the
Schwinger process of $e^{+}e^{-}$ pairs creation is exponentially suppressed.
For $t>t_{\mathrm{ds}}$ the Schwinger process becomes relevant and $e^{+}%
e^{-}$ pairs are created. As shown in Refs.~\refcite{RVX03a},\refcite{RVX02}
the pairs created at radius $r_{0}<r_{\mathrm{ds}}$ oscillate with
ultrarelativistic velocity and partially annihilate into photons. At the same
time the electric field oscillates around zero and the amplitude of such
oscillations decreases in time: in a time of the order of $10^{2}-10^{4}$
$\hbar/m_{e}c^{2}$ the electric field is effectively screened to about the
critical value; more precisely, the average of the electric field
$\mathcal{E}$ over one period of oscillation is $0$, but the average of
$\mathcal{E}^{2}$ is of the order of $\mathcal{E}_{c}^{2}$. As a result an
energy density has been deposited\cite{RV02} on the pairs and the photons
given by
\begin{equation}
\epsilon_{0}\left(  r_{0}\right)  =\frac{1}{8\pi}\left[  \mathcal{E}%
^{2}\left(  r_{0}\right)  -\mathcal{E}_{c}^{2}\right]  =\tfrac{\mathcal{E}%
_{c}^{2}}{8\pi}\left[  \left(  \tfrac{r_{\mathrm{ds}}}{r_{0}}\right)
^{4}-1\right]  . \label{eps0}%
\end{equation}
The pairs and the photons are expected to thermalize\cite{PRX98,RVX03a,RVX02},
to an $e^{+}e^{-}\gamma$ plasma equilibrium configuration:
\begin{equation}
n_{e^{+}}=n_{e^{-}}\simeq n_{\gamma}=n_{0}, \label{n}%
\end{equation}
(where $n_{\bullet}$ is the proper number density of particles of type
$\bullet$), and reach an average temperature $T_{0}$ such that
\begin{equation}
\epsilon\left(  T_{0}\right)  \equiv\epsilon_{\gamma}\left(  T_{0}\right)
+\epsilon_{e^{+}}\left(  T_{0}\right)  +\epsilon_{e^{-}}\left(  T_{0}\right)
=\epsilon_{0}; \label{eq0}%
\end{equation}
here $\epsilon_{\bullet}\left(  T\right)  $ is the equilibrium proper energy
density at temperature $T$ for the species $\bullet$. Then $n_{e^{\pm}}$
($n_{\gamma}$) are given by Fermi (Bose) integrals once the temperature
$T_{0}$ is known.

\section{Plasma's Expansion}

The highly energetic plasma so formed undergoes a relativistic expansion. As
will be shown, the expansion (hydrodynamic) time-scale is much bigger than
both the pair creation and the thermalization time-scales, then the process
can be described as follows: at any time $t_{0}$ it begins to expand a slab of
plasma of thickness $\Delta l=$ $\alpha^{-1}\Delta r$ (as measured by static
observers) produced at radius $r_{0}=r_{0}\left(  t_{0}\right)  $. $\Delta l$
can be chosen very small in comparison with $r_{\mathrm{ds}}$ so that, in
particular, the temperature $T$ is approximately constant in the slab.
Moreover $\Delta l$ has to be much bigger than the quantum length scale
($\sim$ $\hbar/m_{e}c$).

We can follow the expansion of each slab of plasma by using conservation of
energy and number of particles. Note that Eqs.~(\ref{eq0}) and (\ref{n})
provide initial data for the problem of the expansion. We describe the
expansion of a single slab using the following approximations:

\begin{enumerate}
\item  the geometry in which the expansion occurs is Reissner-Nordstr\"{o}m
with line-element given by (\ref{ds}). In particular we will denote by
$\xi^{a}$ the static vector field normalized at unity at spatial infinity, and
by $\left\{  \Sigma_{t}\right\}  _{t}$ the family of space-like hypersurfaces
orthogonal to $\xi^{a}$ ($t$ being the Killing time);

\item  the plasma is assumed to be a neutral perfect fluid characterized by
proper energy density $\epsilon$, proper pressure $p$, proper particle
(electrons, positrons and photons) number density $n$ and $4-$velocity $u^{a}$;

\item  at any instant, electrons, positrons and photons in a single slab are
assumed to be at thermal equilibrium with temperature $T$, possibly different
from slab to slab. The slabs are uncorrelated in the sense that they do not
share energy nor particles. In other words the expansion of each slab is
adiabatic; this will be checked \textit{a posteriori} (see also Ref.~\refcite
{RSWX00});

\item  the thickness $\Delta l=\alpha^{-1}\Delta r$ of a slab as measured by
static observers is constant.
\end{enumerate}

\begin{figure}[ptb]
\begin{center}
\includegraphics[width=8.5cm,clip]{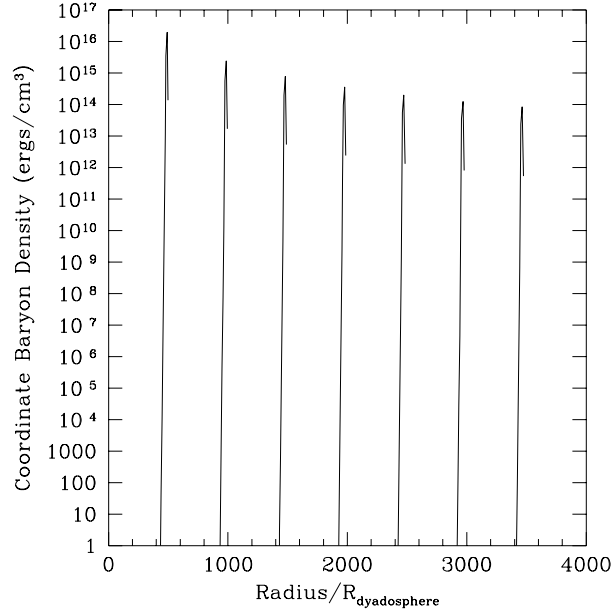}
\end{center}
\caption{A sequence of snapshots of coordinate baryon energy density is shown
from the numerical solution of partial differential continuity equations. This
run correspond to an EMBH of mass $M=10^{3}M_{\odot}$ and charge to mass ratio
$\xi\equiv Q/M=0.1$.}%
\label{slab}%
\end{figure}

The last approximation is justified by the result in Ref.~\refcite{RSWX00},
where it is shown, by numerical integration of the partial differential
continuity equations, that the baryon energy density of an expanding slab of
plasma enriched with nucleons from the remnant of the progenitor star is
localized in a region of constant thickness (see Fig.~\ref{slab}).

\begin{figure}[ptb]
\begin{center}
\includegraphics[width=8.5cm,clip]{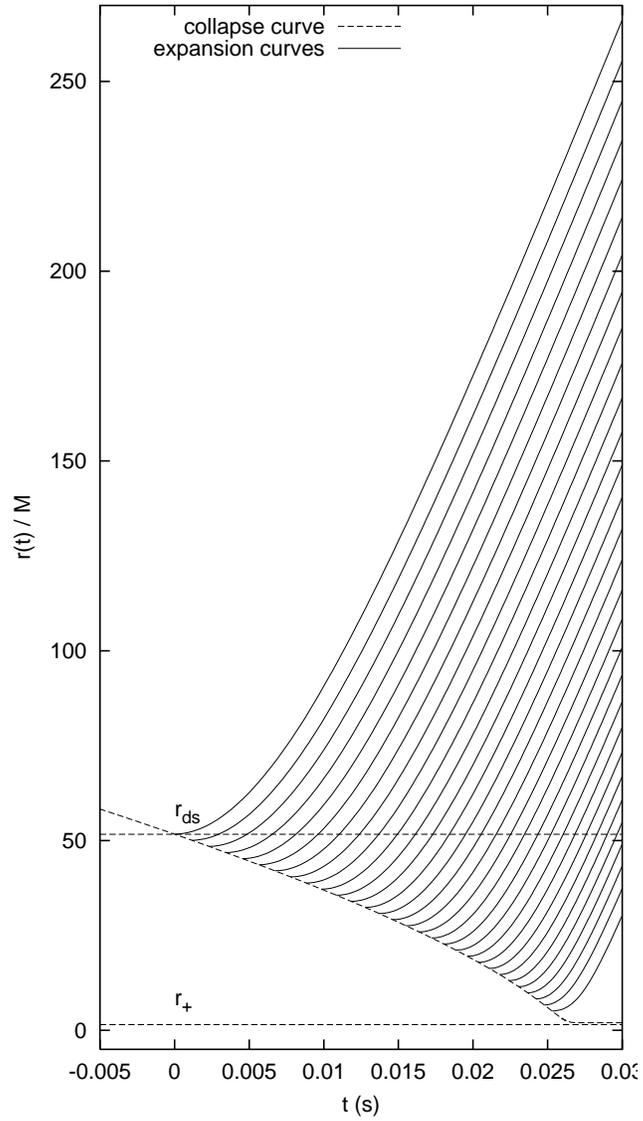}
\end{center}
\caption{Collapse curve of an elettromagnetic star core (dashed line) as
derived by Eq.~(\ref{Motion}) in the case $M=M_{0}=20M_{\odot}$, $Q=0.1M$ and
expansion curves of plasma.}%
\label{expansion}%
\end{figure}

Given the above assumptions, both the energy momentum--tensor $T^{ab}=\left(
\epsilon+p\right)  u^{a}u^{b}+pg^{ab}$ and the electron (positron) --number
current $n_{e}^{a}=n_{e}u^{a}$ are conserved:
\begin{align}
\nabla_{b}T^{ab}  &  =0,\label{Tab}\\
\nabla_{b}n_{e}^{b}  &  =0. \label{nb}%
\end{align}
In particular, using assumption $\left(  4\right)  $ one can reduce the
partial differential continuity equations (\ref{Tab}) and (\ref{nb}) to
ordinary differential equations for the radial coordinate $r$ and the
temperature $T$ of the single slab as functions of time (see Ref.~\refcite
{RSWX00}). The equation of motion of a single slab can be numerically
integrated with initial conditions
\begin{align}
r\left(  t_{0}\right)   &  =r_{0},\\
\left.  \tfrac{dr}{dt}\right|  _{t=t_{0}}  &  =0,\\
T\left(  t_{0}\right)   &  =T_{0}.
\end{align}
The overall motion of the plasma is the superposition of motions of single
shells. The typical plasma expansion curves are shown in Fig.~\ref{expansion}
from the numerical integration of the equations of motion.

The curvature of space--time strongly affects the motion of plasma in the
vicinity of the EMBH horizon and in turn the phenomenology of the GRB. We
discuss these issues in a forthcoming paper.\cite{RVX03}


\begin{thebibliography}{99}
\bibitem{ruffc}D. Christodoulou and R. Ruffini, \textit{Phys.\ Rev.}
\textbf{D4}, 3552 (1971).

\bibitem{DR75}T. Damour, R. Ruffini, \textit{Phys. Rev. Lett.} \textbf{35},
463 (1975).

\bibitem{he35}W. Heisenberg and H. Euler, \textit{Zeits.\ Phys.} \textbf{98},
714 (1935).

\bibitem{s51}J. Schwinger, \textit{Phys.\ Rev.} \textbf{98}, 714 (1951).

\bibitem{RV02}R. Ruffini, L. Vitagliano, \emph{Phys. Lett. B }\textbf{545}
(2002) 233.

\bibitem{PRX98}G. Preparata, R. Ruffini and S.-S. Xue, \textit{A\&A}
\textbf{338}, L87 (1998).

\bibitem{RBCFX01a}R. Ruffini, C. L. Bianco, P. Chardonnet, F. Fraschetti and
S.-S. Xue, \textit{ApJ} \textbf{555}, L107 (2001).

\bibitem{RBCFX01b}R. Ruffini, C. L. Bianco, P. Chardonnet, F. Fraschetti and
S.-S. Xue, \textit{ApJ} \textbf{555}, L113 (2001).

\bibitem{RBCFX01c}R. Ruffini, C. L. Bianco, P. Chardonnet, F. Fraschetti and
S.-S. Xue, \textit{ApJ} \textbf{555}, L117 (2001).

\bibitem{rw75}R. Ruffini and J.R. Wilson, \textit{Phys.\ Rev.} \textbf{D12},
2959 (1975).

\bibitem{s70}V.F. Shvartsman, \textit{Sov.\ Phys.\ JETP} \textbf{33}, 475 (1970).

\bibitem{punsly_book}B. Punsly, \textit{Black Hole Gravitohydromagnetics},
Springer, 2001.

\bibitem{gj69}P. Goldreich and W.H. Julian, \textit{Ap.\ J.} \textbf{157}, 869 (1969).

\bibitem{RMG}R. Ruffini, in \emph{Proceedings of the Ninth Marcel Grossmann
Meeting on General Relativity}, Gurzadyan V.G., Jantzen R.T. \& Ruffini R.
editors, World Scientific, Singapore (2002).

\bibitem{RSWX99}R. Ruffini, J. D. Salmonson, J. R. Wilson and S.-S. Xue,
\textit{A\&A} \textbf{350}, 334 (1999).

\bibitem{RSWX00}R. Ruffini, J. D. Salmonson, J. R. Wilson and S.-S. Xue,
\textit{A\&A} \textbf{359}, 855 (2000).

\bibitem{BRX00}C. L. Bianco, R. Ruffini, S.-S. Xue, \textit{A\&A}
\textbf{368}, 377 (2000).

\bibitem{RVX03}R. Ruffini, L. Vitagliano and S.-S. Xue, (2003) \emph{in
preparation}.

\bibitem{RVX03a}R. Ruffini, L. Vitagliano and S.-S. Xue, \textit{Phys. Lett. }
\textbf{B559}, 12 (2003).

\bibitem{RVX02}R. Ruffini, L. Vitagliano and S.-S. Xue, (2003) \emph{in these
proceedings}.

\bibitem{CRV02}C. Cherubini, R. Ruffini and L. Vitagliano, \textit{Phys. Lett.
} \textbf{B545}, 226 (2002).
\end{thebibliography}
\end{document}